# Curved spacetime-induced control of photonic modes via spatially dependent band structure

JINGXUAN ZHANG, SUTING JU, AND LI-GANG WANG*

*School of Physics, Zhejiang University, Hangzhou 310058, China*
**Corresponding author: lgwang@zju.edu.cn*

**Abstract:** Conventionally, controlling photonic modes require complex artificial structures made of electromagnetic media such as photonic crystal, metamaterial, and waveguide systems. Here, we report a new mechanism for mode control induced solely by curved spacetime, which give rise to a spatially dependent photonic band structure. In this framework, the photonic mode can naturally undergo conversion in the spatial domain. We select two canonical models from general relativity—the Rindler spacetime and the Einstein-Rosen bridge (ERB)—to demonstrate light propagation effects. In Rindler spacetime, a light beam transitions to a diffusive mode for positive acceleration and to a highly collimated propagating mode for negative acceleration. In the ERB, beam transmission is governed by the Schwarzschild radius, which determines the extend of the spatial bandgap. Furthermore, an intriguing tunneling effect is also illustrated. Finally, we propose several feasible experimental methods to verify our theoretical predictions. Our findings elucidate a distinctive formation mechanism of photonic band structure in curved spacetime, enabling precise spatial control of light and the design of photonic devices within a non-Euclidean geometrical framework.

## 1. INTRODUCTION

The control of light propagation through electromagnetic media is a fundamental engineering challenge in modern photonics. This field advanced significantly in 1987 with the first proposal of photonic crystals, which inhibit light transmission within specific frequency ranges by forming photonic bandgaps [1,2]. This proposal sparked the rapid development of photonic crystal materials, expanding from one to three dimensions [3]. Introducing defects into such crystals enables strong light localization [4,5] or efficient guided propagation [6,7]. Artificially engineered metamaterials with tailored refractive indices can also manipulate photonic modes. Negative-index media enhance evanescent waves for super-resolution imaging [8-11], while zero-index media—characterized by infinite effective wavelength and uniform phase distribution [12]—enable many interesting phenomena and potential applications [13]. Exotic phenomena, for instance, invisible tunneling in an optical wormhole [14], the perfect lens and the optical Aharonov-Bohm effect [15], are also expected to be realized in metamaterials. Topological photonics can achieve robust, directional light propagation via protected edge states [16-18], and gradient-index waveguides also facilitate mode conversion [19,20]. Yet these methods typically require complex artificial structures. In this work, inspired by Einstein's general relativity, we introduce a mechanism to control photonic modes in the spatial domain without relying on any electromagnetic medium or material.

Over the past decade, numerous novel optical effects in curved spacetime have attracted significant attention. To simulate in the laboratory the dynamics of light in curved spacetime, a variety of optical analogue systems have been proposed [21-30]. These include electromagnetic fluids with extremely low group velocity [21], radially graded-index metamaterials [22,23], microstructured optical fibers [24], Newton-Schrödinger systems [25-27], and gradient-index lenses or waveguides [28-31]. In parallel, the geometry of curved space itself can be physically constructed to directly investigate intrinsic optical effects [32-47]. Research in this area has explored many phenomena, including the propagation of Gaussian beam on surfaces of revolution (SORs) [32,33], the Wolf effect and spectral switch of partially coherent light fields [34-36], super-resolution imaging [37], spatially accelerating electromagnetic wave packets [38,39], the Gouy phase shift [40], topological edge states in curved waveguide arrays [41], chaos manipulation in curved microcavities [42,43], the light tunneling effect on a hollow curved waveguide [44], as well as studies of Hanbury Brown and Twiss correlations [45], surface plasmon polaritons [46], and photon-sphere modes in curved microcavities [47].

In this work, we investigate the optical consequences of two classic models from general relativity: the Rindler accelerating reference frame and the Einstein-Rosen bridge (ERB). Guided by Einstein's equivalence principle—which asserts the local equivalence between acceleration and spacetime curvature—we first analyze the wave equation for light in Rindler spacetime. We find that an accelerating reference frame transforms propagating plane waves into evanescent waves, creating a spatially dependent photonic band structure that may force a propagating mode into a diffusive mode. Conversely, deceleration facilitates the conversion of evanescent waves into propagating modes, enabling the focusing of a divergent beam into a highly collimated propagating mode. We then examine light propagation on an ERB, a standard model for a wormhole geometry. Here, plane-wave components with nonzero transverse spatial frequencies encounter centrally symmetric potential barriers. Within these barriers, these components become evanescent, tunneling through before reverting to propagating waves. Using these models, we demonstrate the transition between propagating and diffusive modes for a Gaussian beam, as well as the tunneling of beams with different incident angles through these effective spatial photonic bandgaps. Finally, we propose feasible experimental methods to verify our theoretical predictions.

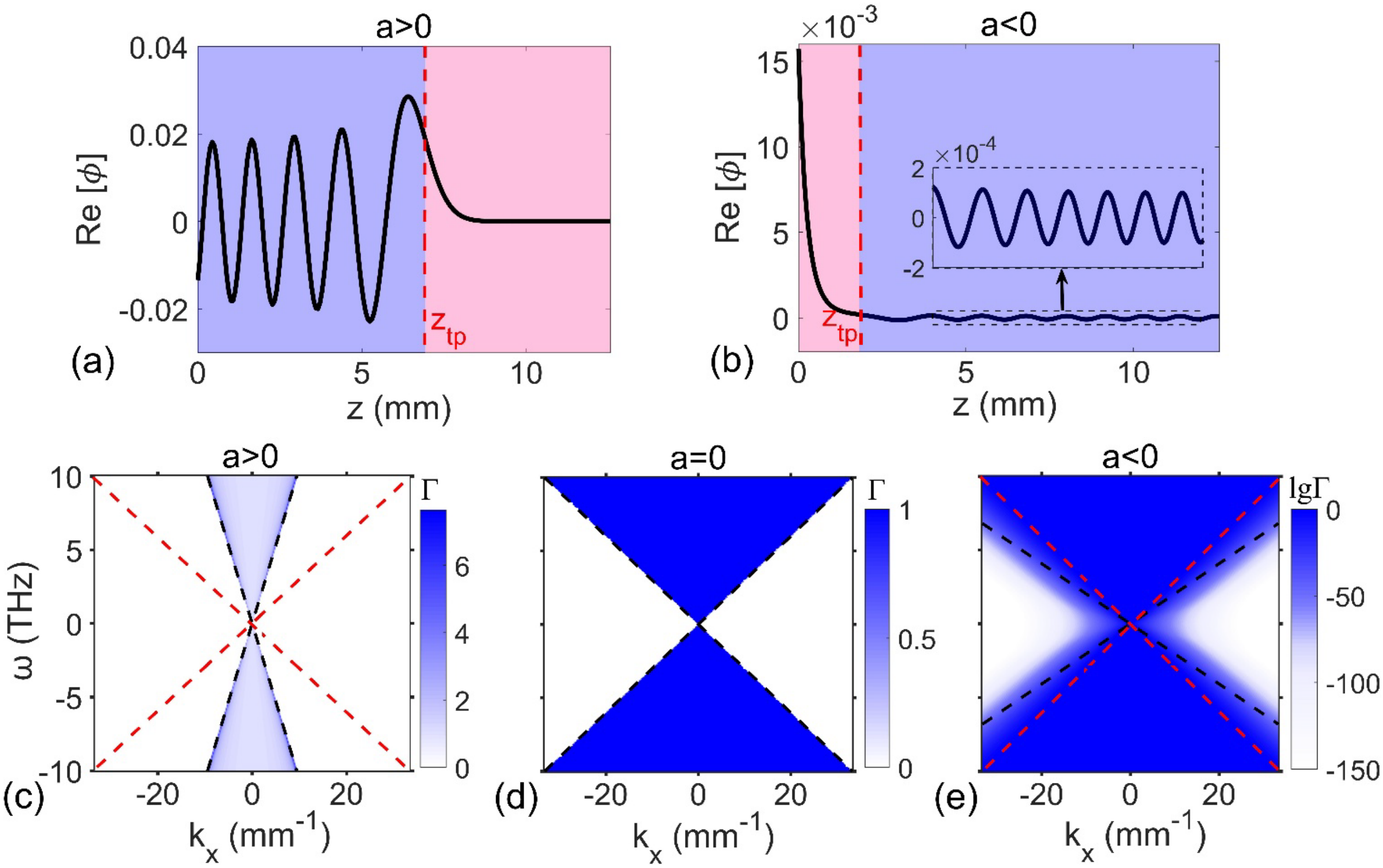

**Fig. 1.** Mode transition and transmission of plane wave component in Rindler spacetime. (a)-(b) Transition between propagating mode (blue region) and evanescent mode (red region) in reference frame with (a) positive and (b) negative acceleration. (c)-(e) Transmitted spectrum at a distance $z_D$, where $\Gamma = \frac{|\phi(z=z_D)|^2}{|\phi(z=0)|^2}$ is defined as transmittance. The red dashedines in (c) and (e) show the lines of light cone in free space. Here $z_D$ is 12.6mm. The acceleration $a/c^2$=100m$^{-1}$ in (a) and (c) while $a/c^2$=−100m$^{-1}$ in (b) and (e). The wavenumber component $k_x$=3.142mm$^{-1}$ in (a) and $k_x$=7.540mm$^{-1}$ in (b), respectively.

## 2. LIGHT DIFFUSION AND FOCUSING IN RINDLER SPACETIME

The 2+1 spacetime metric experienced by an accelerated reference frame is: $ds^2 = e^{\frac{2az}{c^2}}(-c^2dt^2 + dz^2) + dx^2$[48], which is the Rindler metric, with $a$ and $c$ being the acceleration of reference frame and speed of light, respectively. It is necessary to emphasize that, here $z$ and $x$ are coordinates in accelerated reference frame. The metric implies an exponential spatiotemporal expansion or contraction effect of accelerated observer. Light waves in such spacetime satisfies the massless Klein-Gordon equation [49]:

$$\Box\psi = \frac{1}{\sqrt{-g}}\partial_\mu\left(\sqrt{-g}g^{\mu\nu}\partial_\nu\psi\right) = 0, \tag{1}$$

where $\Box$ is the d'Alembert operator, $g$ and $g^{\mu\nu}$ represent the determinant and inverse element of metric tensor matrix $\boldsymbol{g}$. Substituting the above 2+1 Rindler metric into Eq. (1), we have:

$$-\frac{1}{c^2}\frac{\partial^2\psi}{\partial t^2} + \frac{\partial^2\psi}{\partial z^2} + e^{\frac{2az}{c^2}}\frac{\partial^2\psi}{\partial x^2} = 0. \tag{2}$$

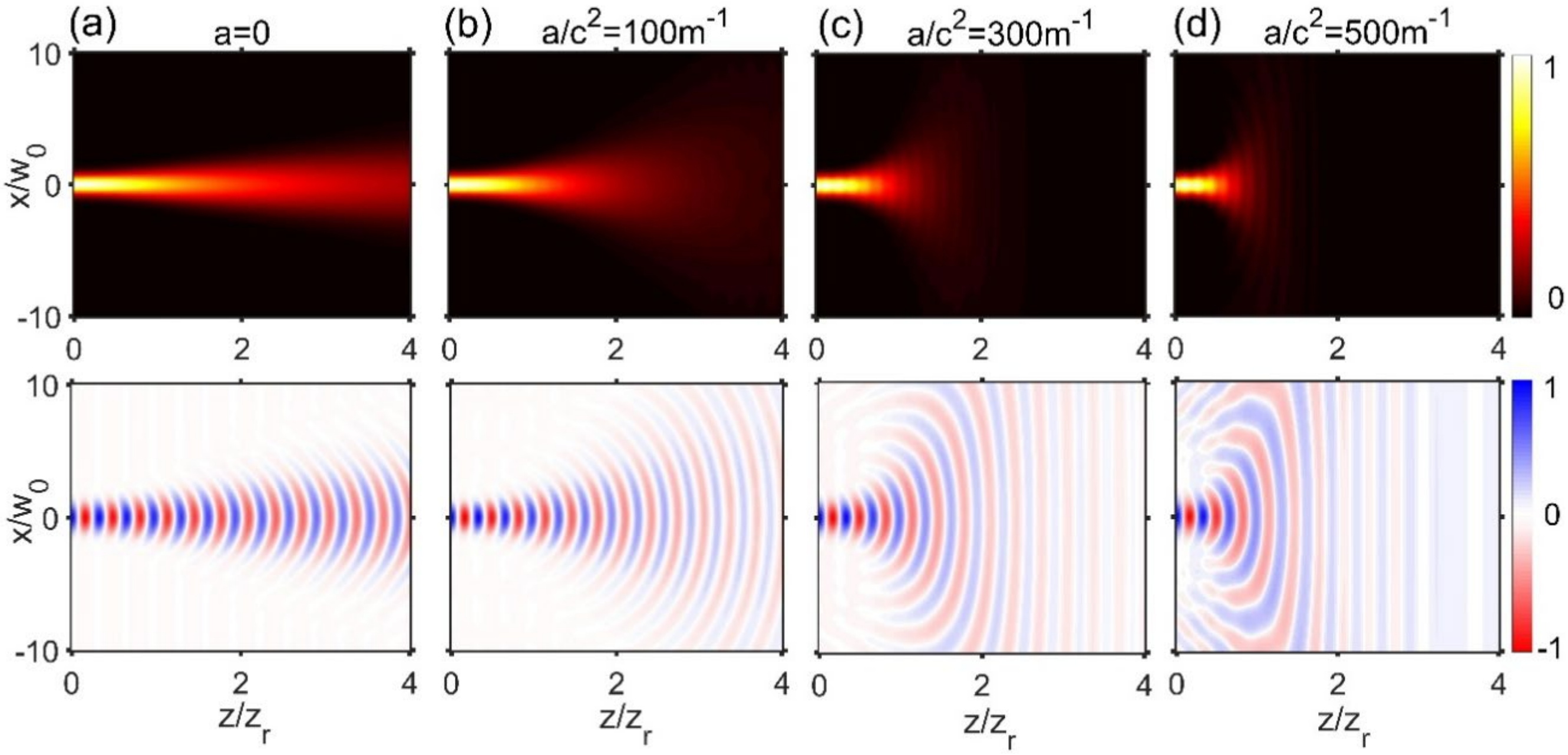

**Fig.2.** Gaussian beam propagation under different values of positive $a$. Here the distance $z$ is reduced by the Rayleigh distances $z_r = \frac{kw_0^2}{2}$ with $w_0$=1mm the initial beam half-width and the wavelength $\lambda$=1mm in the calculation. Here the top row is the intensity distribution $|\psi|^2$ and the bottom row is the evolution of field amplitude Re $[\psi]$ (indicating the phase changes).

Applying variable separation $\psi(t,x,z) = T(t)X(x)\phi(z)$, Eq. (2) can be separated into three ordinary differential equations

$$\frac{d^2T}{dt^2} + k^2c^2T = 0, \tag{3a}$$

$$\frac{d^2X}{dx^2} + k_x^2X = 0, \tag{3b}$$

$$\frac{d^2\phi}{dz^2} + L^2(k_x,z)\phi = 0, \tag{3c}$$

where $L^2(k_x,z) = k^2 - e^{\frac{2az}{c^2}}{k_x}^2$, $k$ is wavenumber of light and $k_x$ is transverse component of $k$. Apparently, the solutions of Eqs. (3a) and (3b) can be directly written as $T = \exp(i\omega t)$ and $X = \exp(ik_xx)$, respectively. For Eq. (3c), the mode transition between propagating mode ($L^2(k_x,z) > 0$) and evanescent mode ($L^2(k_x,z) < 0$) occurs at the zero (also called the turning point (TP)) of $L^2(k_x,z)$ (see Fig. 1(a) and (b)). In the region $z$>0, there are two scenarios that make the position of TP $z_{\text{tp}} = \frac{c^2}{2a}\ln\left(\frac{k^2}{k_x^2}\right)$ positive such that mode transition occurs during light propagation. When $a$>0 and $k_x$<$k$, the initially propagating mode transitions into an evanescent mode. Conversely, when $a$<0 and $k_x$>$k$, the initially evanescent mode transitions into a propagating mode. Therefore, the number of propagating modes decreases/increases when acceleration of reference frame is positive/negative. The exact solution of Eq. (3c) is discussed in Appendix. Now, by fixing a certain time ($t$=const), the final wave function is the combination of all eigenmodes:

$$\psi(x,z) = \int_{-\infty}^{\infty} f(k_x)\phi(k_x,z)\, e^{ik_xx}dk_x, \tag{4}$$

where $f(k_x)$ is the angular spectrum, which can be obtained through the initial wave function at $z$=0 by $f(k_x) = \frac{1}{2\pi\phi(k_x,z=0)}\int_{-\infty}^{\infty}\psi(x,z=0)\, e^{-ik_xx}dx$.

In conventional photonic crystals, a photonic bandgap refers to a range of frequencies (or energies) in which light propagation is prohibited. This is fundamentally an energy (frequency)-domain concept. Here we reveal that under certain curved spacetime backgrounds—such as Rindler spacetime with positive acceleration—light propagation also becomes forbidden for some frequencies with non-zero transverse wavenumber, and this prohibition depends on the spatial propagation distance (see Fig. 1(a)). This spatially dependent inhibition of light propagation is conceptually analogous to a bandgap. To capture this conceptual parallel while avoiding confusion, we introduce the term "spatially dependent photonic bandgap" to describe this effect. In the presence of positive acceleration, the transmitted band—represented by the blue region inside the light cone—becomes narrower, while the forbidden region (white area outside the light cone) expands (see Fig. 1(c)). This clearly demonstrates how the allowed region for propagation is modified in a spatially dependent manner under the influence of acceleration. From the transmitted spectrum in Fig. 1(e), the transmitted band (blue region inside the light cone) becomes wider, while the forbidden band (white region outside the light cone) becomes narrower when negative acceleration is present.

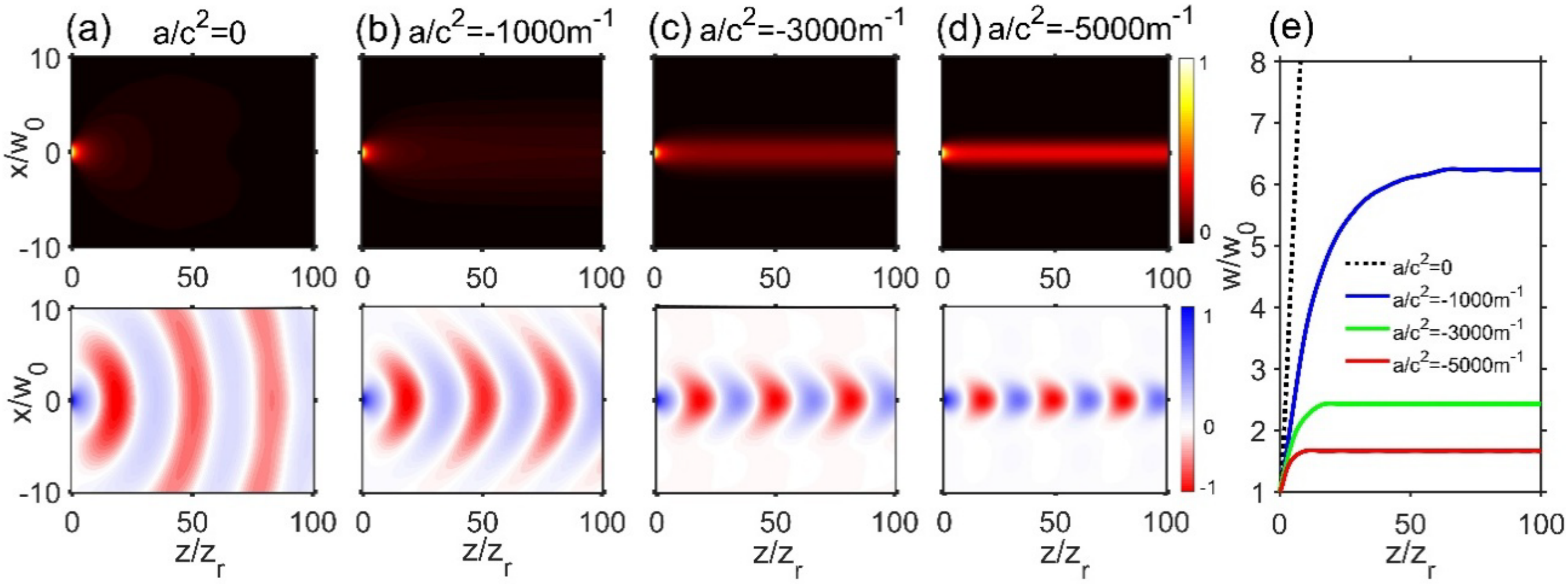

**Fig. 3.** Collimated propagating mode for a divergent beam with $w_0$=0.1mm for negative $a$. (a)-(d) Beam focusing effects under different effective gravity. (e) Evolution of beam width under different accelerations in (a)-(d). Here the top and bottom row in (a)-(d) represent the evolution of intensity $|\psi|^2$ and field amplitude Re $[\psi]$, respectively. The wavelength of light is the same as that in Fig. 2.

To ensure a clear presentation of the physical phenomenon, special attention should be paid to the appropriate selection of the acceleration $a$ and propagation distance $z$. In Rindler metric, the exponential factor $e^{\frac{2az}{c^2}}$ governs the observed optical effects, depending on both $a$ and $z$. For a fixed and modest distance, a very large acceleration is required to make this factor sufficiently deviate from unity and produce a discernible effect. Conversely, with a modest acceleration, an impractically large propagation distance would be needed to achieve the same factor. For instance, the acceleration in Fig. 1(a) is $a$=100$c^2$=9 × $10^{18}$m/s$^2$ over a distance of $z$=0.0126m. To achieve the same exponential factor with a much smaller acceleration of $a$=1×$10^3$m/s$^2$, the required propagation distance would be $z$=1.134×$10^{14}$m (approximately 0.012 light-years). At such distances, the natural divergence of a Gaussian beam would disrupt the effect. Therefore, in our example, we fix the propagation distance at a practical $z$=0.0126m (four times the Rayleigh distance) and employ correspondingly large accelerations to clearly demonstrate the optical signatures of Rindler spacetime.

Now, we consider the propagation of a Gaussian beam: $\psi(x, z=0) = \exp(-x^2/w_0^2)$ in Rindler spacetime, where $w_0$ is initial beam half-width. When the reference frame is accelerating, most high-frequency components of light wave transition from propagating waves into evanescent waves, leading to energy diffusion in the propagation direction (see Fig. 2). Only the mode of $k_x$=0 can propagate over a distance. This effect can be explained by Einstein's relativity: the dynamical effect of acceleration is locally equivalent to gravity acting in the opposite direction. In Fig. 2, acceleration is along +$z$ direction, which generates an effective gravity along −$z$ direction acting on Gaussian beam, causing a squeezing effect of the beam in $z$ direction. This squeezing effect makes the beam wider and produces interference fringes because some portion of light turns back from the TP. Against this curved spacetime background, the forward propagation of light beam components with nonzero transverse wavenumber is prohibited, giving rise to a spatial photonic bandgap. As $a$ increases, the energy diffusion occurs faster along the propagation direction, indicating the spatial bandgap becomes wider. When reference frame is decelerating for negative values of $a$, most evanescent waves can transition into propagating waves, which recovers more spatial information over a distance, leading to beam focusing (see Fig. 3). Similarly, when the acceleration is along −$z$ direction, the corresponding effective gravity is along +$z$ direction; consequently, the divergent beam ($w_0$=0.1mm) appears to be straightened in the $z$ direction and focuses to a beam with constant width, as shown in Fig. 3 (e).

Since the values of acceleration in the above calculation are extremely huge (in the unit of $c^2$), which makes them hard to realize in laboratory, we propose a possible experiment based on the dynamical equivalence between the 2+1 Rindler spacetime and a curved surface embedded in three-dimensional space. By considering the null geodesic (trajectory of massless particle, like photon): $ds^2 = 0$, the 2+1Rindler metric can be written as a Fermat metric [43]:

$$ds_F^2 = dz^2 + e^{-\frac{2az}{c^2}}dx^2. \quad (5)$$

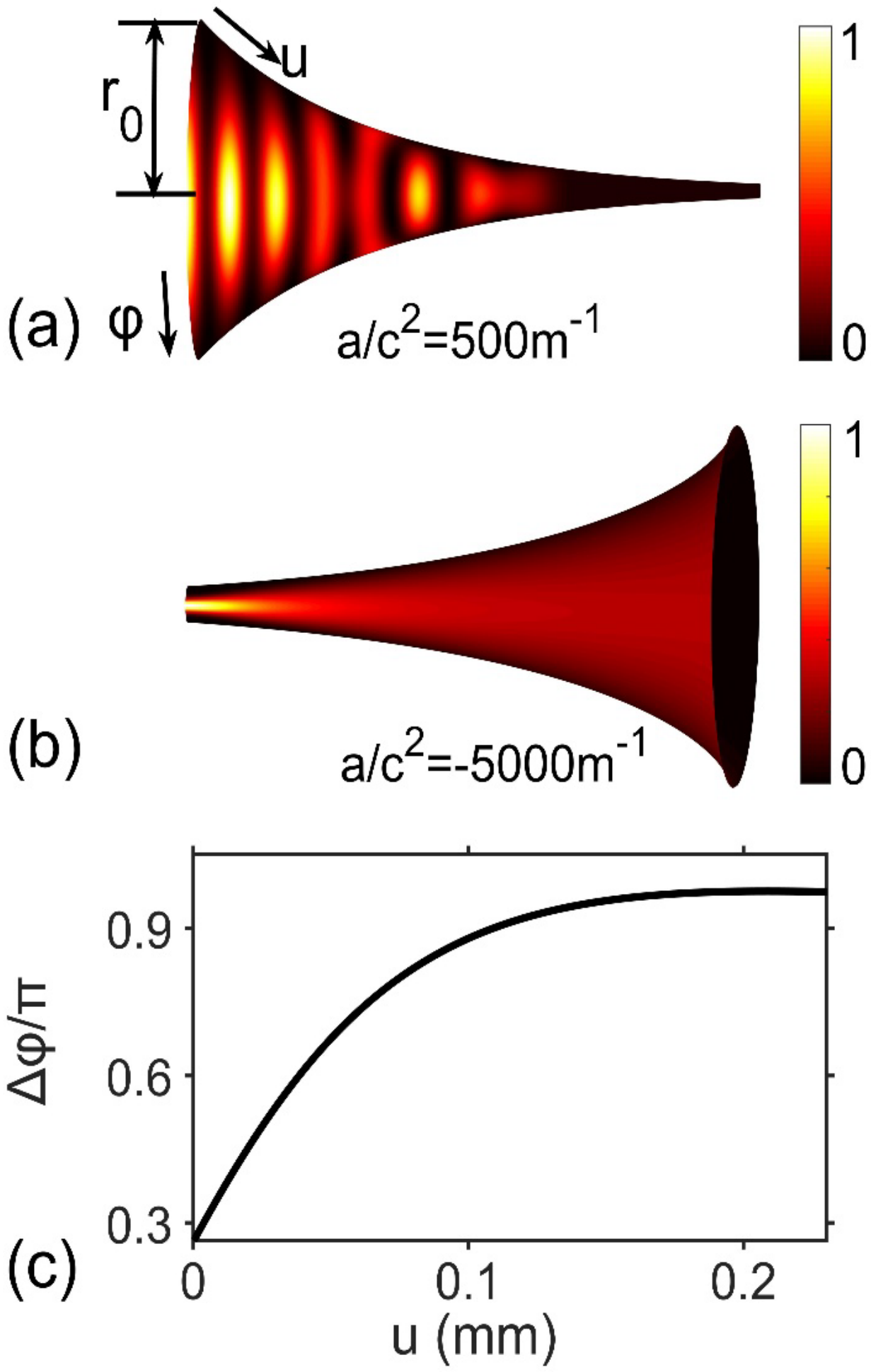

**Fig. 4.** SORs that equivalent to Rindler spacetime with positive and negative accelerations. (a) Beam diffusion and (b) angular focusing. (c) evolution of angular beam width in (b).

Equation (5) is an optical metric for light propagation in the Rindler spacetime, which is mathematically equivalent to the surface of revolution (SOR) with the metric $ds^2 = du^2 + r_0^2 e^{-\frac{2au}{c^2}}d\varphi^2$. This equivalence can be established by the coordinate transformation: $z = u, x = r_0\varphi$, where $u$ is the proper length along the longitudinal line of SOR, $r_0$ is initial rotational radius, and $\varphi$ is rotational angle, as shown in Fig. 4. The accelerating/decelerating effect can then be demonstrated through the variation of rotational radius of SOR: $r(u) = r_0 e^{-\frac{au}{c^2}}$. Thus, the extremely huge acceleration can be experimentally realized by constructing SORs whose rotational radius varies as $e^{-\frac{au}{c^2}}$. We also illustrate this equivalence in Fig. (4). Figure 4(a) shows that the Gaussian beam cannot propagate on SOR with positive $a$. For SOR with negative $a$, the focusing effect (a constant beam width during propagation) shown in Fig. 3 can be realized by evaluating the angular width of beam (see Fig. 4(c)) since $x$ coordinate in Rindler spacetime is replaced by the rotational angle $\varphi$ on SOR.

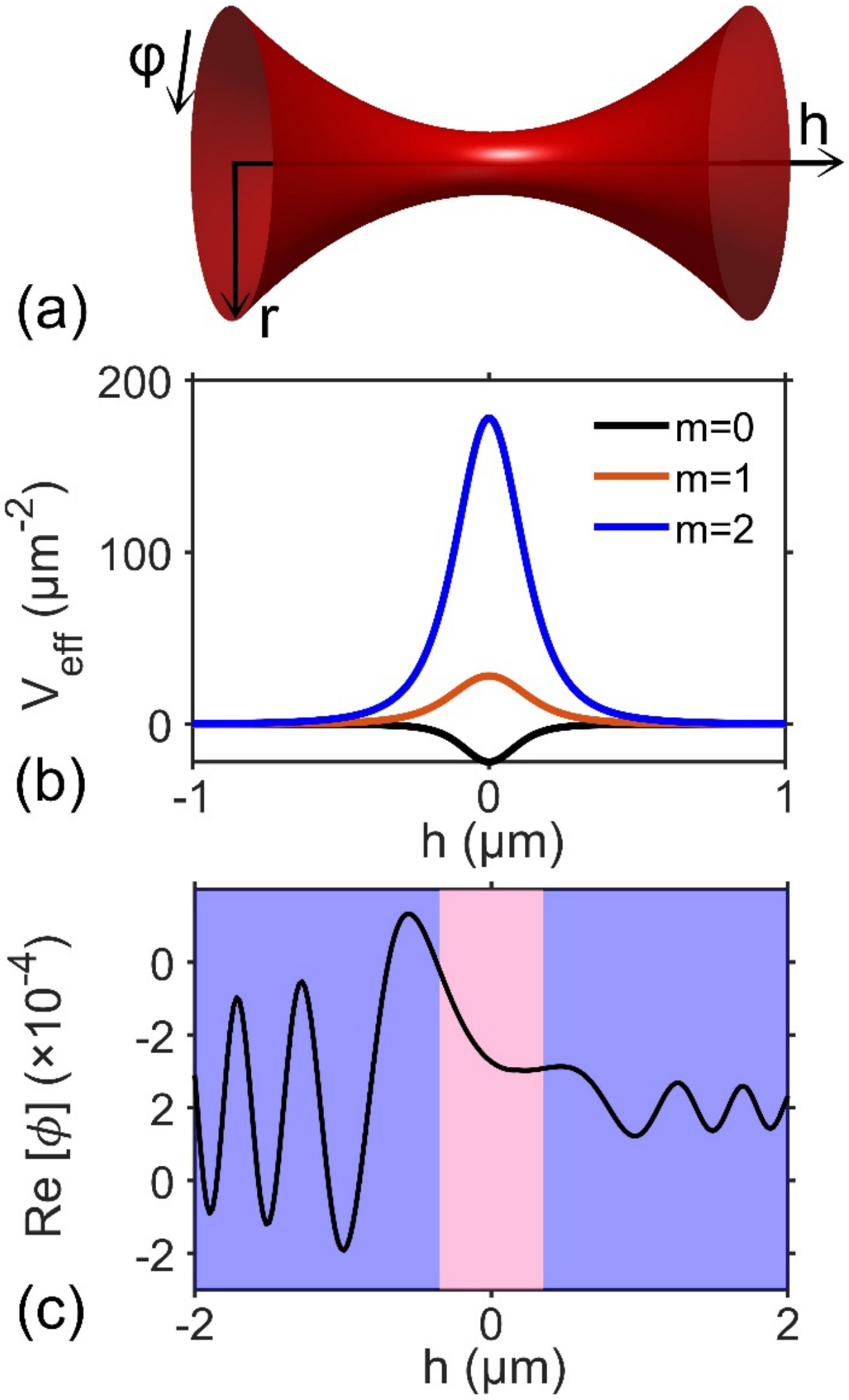

**Fig. 5.** Effective potential barrier and mode transition on an ERB. (a) Schematic illustration of an ERB. (b) Effective potential well or barrier for plane-wave components with different angular discrete number $m$. (c) Tunneling through a barrier near the bottleneck of an ERB. The Schwarzschild radius $r_s$=0.1μm in (b) and $r_s$=1 μm in (c). The angular discrete number in (c) is $m$=16.

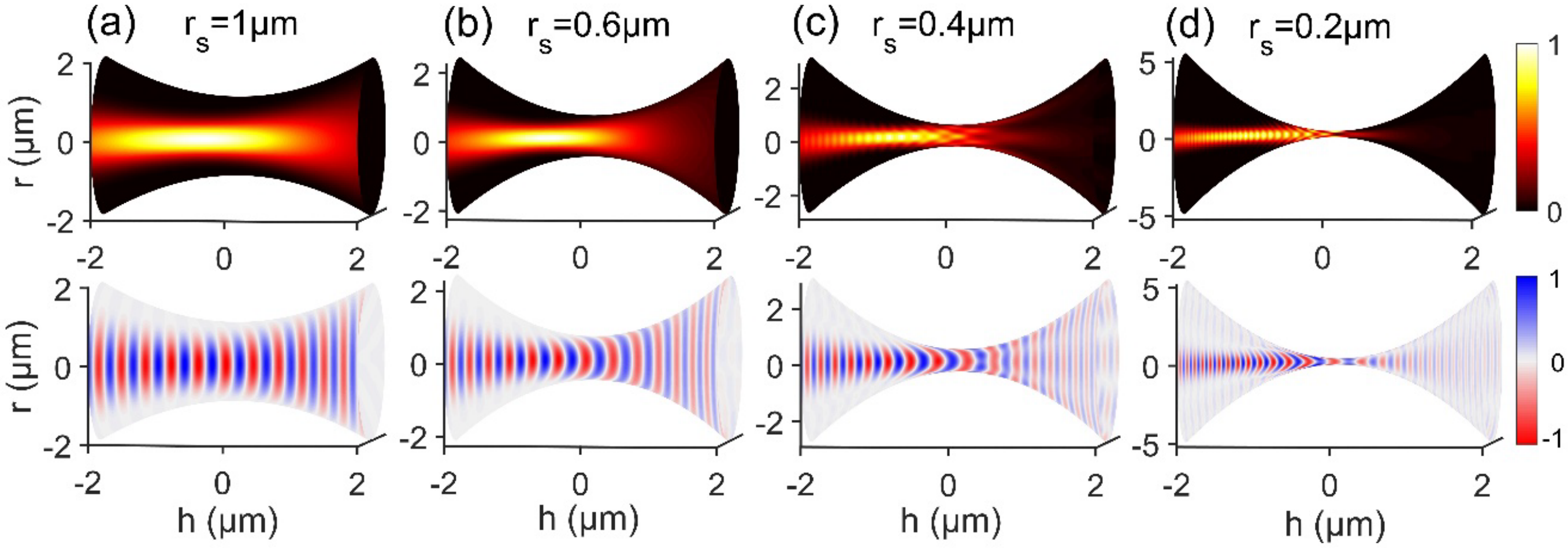

**Fig. 6.** Transmission of a Gaussian beam on ERBs. (a)-(b) The bottlenecks of ERBs are traversable for light beam. (c)-(d) Diffusive mode near bottleneck when the space is strongly curved. We show the distribution of intensity $|\psi|^2$ in top row and field amplitude Re $[\psi]$ in bottom row. The wavelength λ=405nm and beam half-width $w_0$=1μm.

## 3. SPATIAL BANDGAP AND PHOTONIC TUNNELING ON EINSTEIN-ROSEN BRIDGE

In this section, we introduce the Einstein-Rosen bridge (ERB) as a typical model of space curved by a wormhole. First, the two-dimensional Schwarzschild metric can be written as $\mathrm{d}s^2=\left(1-\frac{r_s}{r}\right)^{-1}dr^2+r^2d\varphi^2$ [50], where $r$ is the rotational radius, $\varphi$ is the rotational angle and $r_s$ is the Schwarzschild radius representing the radius at the bottleneck of ERB, as seen in Fig. 5(a). Then, by introducing the following coordinate transformation: $h=\pm2\sqrt{r_s(r-r_s)}$, the metric can be constructed to an ERB: $ds^2=f(h)dh^2+r_s^2f^2(h)d\varphi^2$, where $f(h)=\frac{h^2}{4r_s^2}+1$ and $h$ is a Cartesian coordinate (the height of ERB). The time-harmonic scalar electromagnetic wave confined on such curved surface then satisfies the equation [33]:

$$\Delta_g\psi+(k^2+H^2-K)\psi=0, \tag{6}$$

where $H$ and $K$ are the mean (extrinsic) and Gaussian (intrinsic) curvatures of the curved surface, respectively. Substituting the ansatz $\psi(\varphi,h)=r(h)^{-\frac{1}{2}}\exp(im\varphi)\phi(h)$ into Eq. (6), we obtain

$$\frac{d^2\phi}{dh^2}+L^2(h,\,m)\phi=0, \tag{7}$$

where $L^2(h,\,m)=\left(\frac{h^2}{4r_s^2}+1\right)\left\{k^2-\frac{m^2}{r_s^2}\left(\frac{h^2}{4r_s^2}+1\right)^{-2}+\left(\frac{h^2}{4r_s^2}+1\right)^{-2}\left[\frac{1}{16r_s^2}+\frac{3}{2(h^2+4r_s^2)}\right]\right\}$, $m$ is angular discrete number. We can observe that in $L^2(h,\,m)$, apart from $k^2$, the other terms can be regarded as an effective potential $V_{\mathrm{eff}}(h,m)=\frac{m^2}{r_s^2}\left(\frac{h^2}{4r_s^2}+1\right)^{-2}-\left(\frac{h^2}{4r_s^2}+1\right)^{-2}\left[\frac{1}{16r_s^2}+\frac{3}{2(h^2+4r_s^2)}\right]$, which forms a centrally symmetric potential well when $m$=0 and becomes a barrier when $m\neq0$ as shown in Fig. 5(b).

Figure 6 shows the propagation of a Gaussian beam on ERBs with different $r_s$. The light beam can propagate to the other side of ERB through the bottleneck when $r_s$ is large, as shown in Fig.6(a) and (b), since the local spatial curvature is small and the potential barrier varies gently near the bottleneck. However, when $r_s$ is small enough, most modes with high spatial frequency are scattered by a sharp barrier near the bottleneck and only a small amount of energy with low transverse spatial frequencies can transmit to another side of the ERB, while the self-interference pattern of beam appears in the incident side of the ERB, as seen Fig.6(c) and (d). In this sense, curved space opens a spatial photonic-bandgap that tends to inhibit light propagation.

Moreover, we consider the tunneling effect under different incident angles. The equivalence between incident angle $\theta$ and angular discrete number $m$ is $\theta=\arcsin\left(\frac{m}{kr_i}\right)$, with $r_i$ being the radius of the latitude line, which is also the location of the initial beam on the ERB. For small incident angles (or $m$), most of beam energy transmits through the throat and reaches the other edge of ERB (see Fig. 7(a)). When incident angle increases, the potential barrier appears and broadens (see Fig. 7(b)-(d)), and most mode components of beam gradually transition from the propagating mode (before the barrier) to evanescent mode (inside the barrier) then back to propagating mode (after the barrier) with some energy loss on other side. The transmittance of beam decays rapidly at large $m$ and beam appears to be reflected by the edge of potential barrier, after which it propagates along geodesic on ERB (see Fig. 7(d)). Moreover, the range of incident angle $\theta$ allowing beam to transmit through the ERB narrows as $r_s$ decreases. This photonic tunneling effect can be exploited for mode control of light in spatial domain.

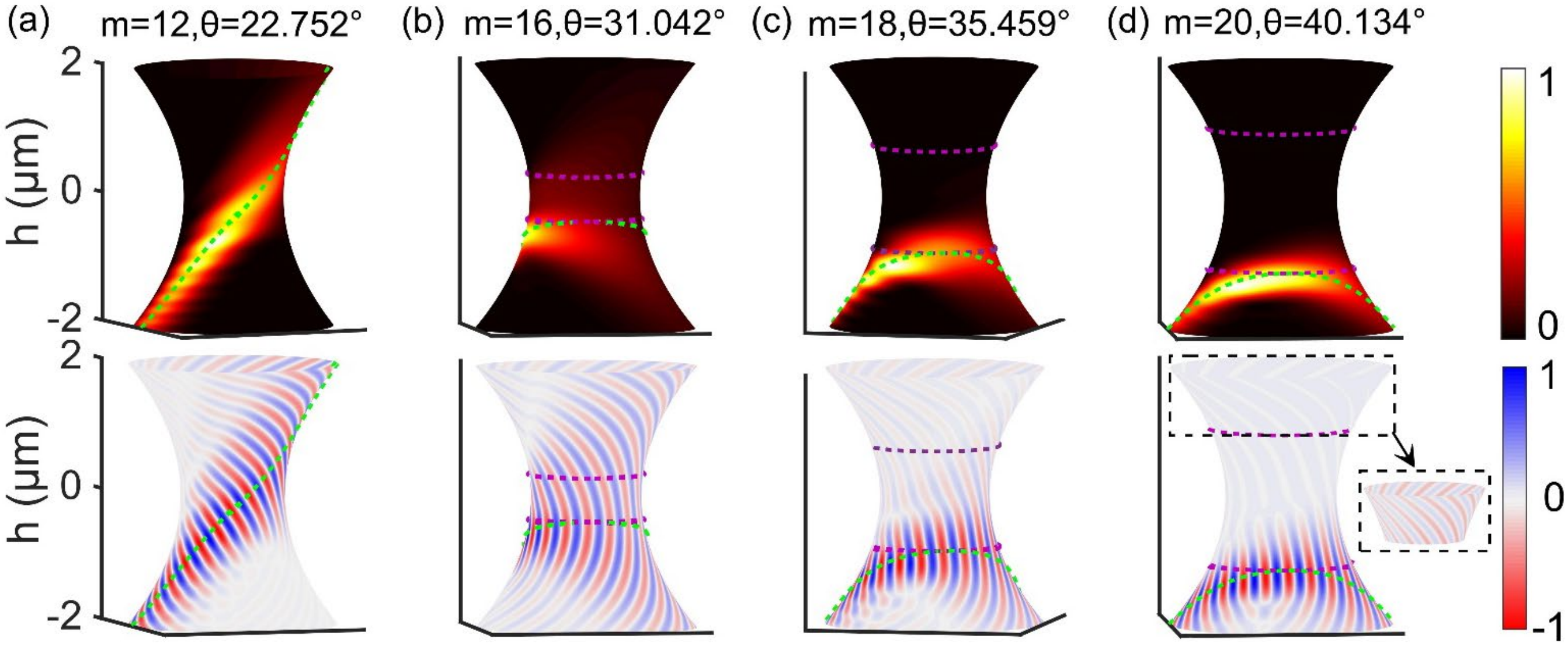

**Fig. 7.** Propagating and tunneling effect for beams with different incident angles. Beam tunnelling through spatial photonic-bandgap (indicated by the region between two purple-dotted lines) on the ERB with $r_s$=1μm. The green-dotted lines are geodesic trajectories of the propagating axis of light beams. The insert near (d) is zoom-in amplitude distribution after tunnelling the spatial bandgap. Here, the first row reveals the intensity distribution $|\psi|^2$ while the second row indicates the evolution of field amplitude Re $[\psi]$. The wavelength and beam half-width are the same as those in Fig. 6.

Experimentally, an ERB structure can be fabricated through a curved waveguide using direct laser writing technology. Researchers in Ref. [44] reported a similar nanostructure by constructing a paraboloid-shaped three-dimensional hollow waveguide and observed optical tunneling phenomenon at the bottleneck of waveguide as well. The rotational radius at bottleneck in Ref. [44] is 4.4μm, which is on the scale of $r_s$ in our work. Other experimental candidates may also be possible, including a wormhole structure fabricated from uniform transparent resin using 3D printing technique [51], a surface with constant negative Gaussian curvature constructed by 3D printed curved waveguide [52] and a negatively curved surface fabricated from bulk aluminum and covered with high index immersion liquid [33].

## 4. CONCLUSION

In summary, we have investigated the transition of photonic modes in spatially dependent band structure, focusing on Rindler spacetime and Einstein-Rosen bridge. In Rindler spacetime, a light beam transitions to diffusive mode during propagation in an accelerating reference frame. Conversely, under deceleration, the beam reverts to a collimated propagating mode. We propose to verify these effects experimentally through a surface of revolution that is equivalent to Rindler spacetime. For the Einstein-Rosen bridge, we show that the beam transmission across the bridge depends on the scale of Schwarzschild radius $r_s$, which determines the extent of a spatial bandgap. By examining light propagation at various incident angles, we visualize a tunneling effect near the spatial bandgap: a portion of the beam energy is reflected, while the remainder tunnels through the gap and recovers to a propagating state.

These optical effects can be experimentally realized by fabricating the corresponding surfaces of revolution using curved thin-waveguide structures and 3D printing techniques. Crucially, all effects demonstrated in this work arise purely from the geometry of curved spacetime, without recourse to complex artificial media or materials. Our results provide a novel perspective for controlling photonic modes in the spatial domain. We believe these results hold significant potential for applications in mode conversion, spatial mode filtering, and the broader field of photonics in curved spacetimes.

## APPENDIX: SOLUTIONS OF EQUATION 3(C) AND EQUATION (7)

In this appendix, we show how to solve Eq.(3c) and Eq. (7). For a specific mode $k_x$, Eq. (3c) can be generally solved by Wentzel-Kramers-Brillouin (WKB) approximation [53]. By substituting the ansatz $\phi(z) = \xi(z)e^{iS(z)}$ into Eq. (3c) and seperating the real and imaginary part, we have

$$\frac{d^2\xi}{dz^2} - \xi\left(\frac{dS}{dz}\right)^2 + L^2\xi = 0, \quad \text{(A1)}$$

$$2\frac{d\xi}{dz}\frac{dS}{dz} + \xi\frac{d^2S}{dz^2} = 0, \quad \text{(A2)}$$

By assuming that $\xi(z)$ is slowly varying, the term $\frac{d^2\xi}{dz^2}$ in Eq. (A1) can be neglected. Then $S(z)$ and $\xi(z)$ can be obtained via the relations $S(z) = \int_0^z \sqrt{L^2(z')}dz'$ and $\xi(z) = [L^2(z)]^{-\frac{1}{4}}$. For simplicity, we define the function $\Lambda_\pm(z_1, z_2) = \int_{z_1}^{z_2} \sqrt{\pm L^2(z')}dz'$. Thus, the solution of Eq. (3c) in the WKB approximation is given by

$$\phi_{\mathrm{WKB}}(z) = [L^2(z)]^{-\frac{1}{4}}\exp[i\Lambda_+(0, z)]. \quad \text{(A3)}$$

However, the WKB solution collapses near TP since Eq. (A3) will diverge. We resort to the modified Airy function (MAF) method to solve Eq. (3c) when TP presents. The solution of Eq. (3c) can be written as a variation of Airy function: $\phi(z) = F(z)Ai[\sigma(z)]$, then Eq. (3c) evolves to

$$F''(z)Ai[\sigma(z)] + Ai'[\sigma(z)][2F'(Z)\sigma'(z) + F(z)\sigma''(z)]$$
$$+Ai[\sigma(z)]F(z)\{\sigma(z)[\sigma'(z)]^2 + L^2(z)\} = 0. \quad \text{(A4)}$$

For slowly varying $F(z)$, the term $F''(z)Ai[\sigma(z)]$ can be neglected. Since $Ai'[\sigma(z)]$ and $Ai[\sigma(z)]$ are linearly independent, Eq. (A4) can be separated by

$$2F'(Z)\sigma'(z) + F(z)\sigma''(z) = 0, \quad \text{(A5)}$$

$$\sigma(z)[\sigma'(z)]^2 + L^2(z) = 0. \quad \text{(A6)}$$

Therefore, the solutions of Eqs. (A5) and (A6) are, respectively, given by

$$\sigma(z) = \begin{cases} -\left|\frac{3}{2}\Lambda_+(z_{\mathrm{tp}}, z)\right|^{\frac{2}{3}}, & \text{if } L^2(z) > 0, \\ \left|\frac{3}{2}\int_{z_{\mathrm{tp}}}^{z}\Lambda_-(z_{\mathrm{tp}}, z)\right|^{\frac{2}{3}}, & \text{if } L^2(z) < 0, \end{cases} \quad \text{(A7)}$$

$$F(z) = \begin{cases} [L^2(z)]^{-\frac{1}{4}}\left|\frac{3}{2}\Lambda_+(z_{\mathrm{tp}}, z)\right|^{\frac{1}{6}}, \text{if } L^2(z) > 0, \\ [-L^2(z)]^{-\frac{1}{4}}\left|\frac{3}{2}\Lambda_-(z_{\mathrm{tp}}, z)\right|^{\frac{1}{6}}, \text{if } L^2(z) < 0. \end{cases} \quad \text{(A8)}$$

Here $z_{\mathrm{tp}} = \frac{c^2}{2a}\ln\left(\frac{k^2}{k_x^2}\right)$ is the position of TP. The lower limit of integral is chosen to be $z_{\mathrm{tp}}$ so that the soluiton is exact enough when $L^2(z)$ is linear in $z$ [53]. Another solution of Eq. (3c) in MAF method is variation of Airy function of the second kind: $\phi(z) = G(z)Bi[\sigma(z)]$. Similar, by substituting this solution into Eq. (3c), we have $G(z)=F(z)$. Therefore, the final solution of Eq. (3c) is a combination of Airy functions of the first and second kinds. When $L^2(z)>0$, the solution is

$$\phi_{\mathrm{MAF}}(z) = [L^2(z)]^{-\frac{1}{4}}\left|\frac{3}{2}\Lambda_+(z_{\mathrm{tp}}, z)\right|^{\frac{1}{6}}$$
$$\times\left\{\alpha Ai\left[-\left|\frac{3}{2}\Lambda_+(z_{\mathrm{tp}}, z)\right|^{\frac{2}{3}}\right] + \beta Bi\left[-\left|\frac{3}{2}\Lambda_+(z_{\mathrm{tp}}, z)\right|^{\frac{2}{3}}\right]\right\}. \quad \text{(A9)}$$

When $L^2(z)<0$, the solution is

$$\phi_{\mathrm{MAF}}(z) = [-L^2(z)]^{-\frac{1}{4}}\left|\frac{3}{2}\Lambda_-(z_{\mathrm{tp}}, z)\right|^{\frac{1}{6}}$$
$$\times\left\{\alpha Ai\left[\left|\frac{3}{2}\Lambda_-(z_{\mathrm{tp}}, z)\right|^{\frac{2}{3}}\right] + \beta Bi\left[\left|\frac{3}{2}\Lambda_-(z_{\mathrm{tp}}, z)\right|^{\frac{2}{3}}\right]\right\}, \quad \text{(A10)}$$

There are asymptotic forms for Airy functions $Ai(q)$ and $Bi(q)$, which are given by

$$Ai(q) \sim \frac{1}{2\sqrt{\pi}}q^{-\frac{1}{4}}e^{-\frac{2}{3}q^{\frac{3}{2}}}, \quad Bi(q) \sim \frac{1}{\sqrt{\pi}}q^{-\frac{1}{4}}e^{\frac{2}{3}q^{\frac{3}{2}}}, \quad q \to \infty,$$

$$Ai(q)\sim\frac{1}{\sqrt{\pi}}(-q)^{-\frac{1}{4}}\sin\left[\frac{2}{3}(-q)^{\frac{3}{2}}+\frac{\pi}{4}\right],$$

$$Bi(q)\sim\frac{1}{\sqrt{\pi}}(-q)^{-\frac{1}{4}}\cos\left[\frac{2}{3}(-q)^{\frac{3}{2}}+\frac{\pi}{4}\right], q\rightarrow-\infty. \quad \text{(A11)}$$

These will be frequently used to determine coefficients $\alpha$ and $\beta$. If the acceleration of reference frame $a$>0, the function $L^2(z)$<0 in the region $z$>$z_{\rm tp}$. Therefore, for $z$ being far large than $z_{\rm tp}$, the asymptotic form of Eq. (A10) is expressed as

$$\phi_{\rm MAF}(z)=[-L^2(z)]^{-\frac{1}{4}}\left\{\frac{\alpha}{2\sqrt{\pi}}\exp[-\Lambda_-(z_{\rm tp},z)]+\frac{\beta}{\sqrt{\pi}}\exp[\Lambda_-(z_{\rm tp},z)]\right\}. \quad \text{(A12)}$$

Since the region with $L^2(z)$<0 is semi-infinite ($z$>$z_{\rm tp}$) and $L^2(z)$ exponentially diverges as $z$ goes to infinity, only the exponential decay solution exists in this region, which leads to $\beta$=0. Therefore, when $a$>0, the exact MAF solution becomes

$$\phi_{\rm MAF}(z)=\alpha[L^2(z)]^{-\frac{1}{4}}\left|\frac{3}{2}\Lambda_+(z_{\rm tp},z)\right|^{\frac{1}{6}}\times Ai\left[-\left|\frac{3}{2}\Lambda_+(z_{\rm tp},z)\right|^{\frac{2}{3}}\right], \text{if } L^2(z)>0,$$

$$\phi_{\rm MAF}(z)=\alpha[-L^2(z)]^{-\frac{1}{4}}\left|\frac{3}{2}\Lambda_-(z_{\rm tp},z)\right|^{\frac{1}{6}}\times Ai\left[\left|\frac{3}{2}\Lambda_-(z_{\rm tp},z)\right|^{\frac{2}{3}}\right], \text{if } L^2(z)<0. \quad \text{(A13)}$$

The MAF solution in Eq. (A13) should equal to the WKB solution in the region with $z$<<$z_{\rm tp}$, i.e.,

$$\phi_{\rm MAF}=\phi_{\rm WKB}=A[L^2(z)]^{-\frac{1}{4}}\exp[i\Lambda_+(0,z)]+B[L^2(z)]^{-\frac{1}{4}}\exp[-i\Lambda_+(0,z)], \quad \text{(A14)}$$

where $A$ and $B$ represent the incident and reflecting coefficients, respectively. Utilizing the asymptotic form of Eq. (A11), we obtain $\alpha=-2\sqrt{\pi}iAe^{i\left[\Lambda_+(0,z_{\rm tp})+\frac{\pi}{4}\right]}$, where $A$ can be simply taken as 1 in calculation. Similarly, when the acceleration is negative ($a$<0), we have the following relation, $\alpha=i\beta$ and $\beta=\sqrt{\pi}e^{-\Lambda_-(0,z_{\rm tp})}$.

The WKB solutions of Eq. (7) is similar to that in the section 2. However, there may exist two TPs on such surface, then the MAF solutions are:

$$\phi_{\delta,\rm MAF}(h,\varphi)=[L^2(h)]^{-\frac{1}{4}}\left|\frac{3}{2}\Lambda_+(h_\delta,h)\right|^{\frac{1}{6}}\times\left\{\alpha_\delta Ai\left[-\left|\frac{3}{2}\Lambda_+(h_\delta,h)\right|^{\frac{2}{3}}\right]+\beta_\delta Bi\left[-\left|\frac{3}{2}\Lambda_+(h_\delta,h)\right|^{\frac{2}{3}}\right]\right\} \quad \text{(A15)}$$

if $L^2(z)$>0, otherwise

$$\phi_{\delta,\rm MAF}(h,\varphi)=[-L^2(h)]^{-\frac{1}{4}}\left|\frac{3}{2}\Lambda_-(h_\delta,h)\right|^{\frac{1}{6}}\times\left\{\alpha_\delta Ai\left[\left|\frac{3}{2}\Lambda_-(h_\delta,h)\right|^{\frac{2}{3}}\right]+\beta_\delta Bi\left[\left|\frac{3}{2}\Lambda_-(h_\delta,h)\right|^{\frac{2}{3}}\right]\right\} \quad \text{(A16)}$$

if $L^2(z)$<0. $\delta$=1,2 indicate TP1 and TP2, respectively, $h_1$ and $h_2$ are positions of two TPs. Here $\phi_{1,\rm MAF}$ is valid near TP1 and collpse near TP2, and vice versa. For simplicity, we use $\phi_{1,\rm MAF}$ when $h$≤0 and $\phi_{2,\rm MAF}$ when $h$>0. Next, the coefficients $\alpha_1$, $\beta_1$, $\alpha_2$ and $\beta_2$ are in need to be determined. In the middle evanescent region ($L^2$<0 and faraway from two TPs), $\phi_{1,\rm MAF}$ and $\phi_{2,\rm MAF}$ should be connected via $\phi_{1,\rm MAF}=\phi_{2,\rm MAF}\phi_{1,\rm MAF}=\phi_{2,\rm MAF}$ . We use the asymbotic form of Airy funtions to rewrite this relation as

$$\frac{\alpha_1}{2\sqrt{\pi}}\exp[-\Lambda_-(h_1,h)]+\frac{\beta_1}{\sqrt{\pi}}\exp[\Lambda_-(h_1,h)]=\frac{\alpha_2}{2\sqrt{\pi}}\exp[-\Lambda_-(h,h_2)]+\frac{\beta_2}{\sqrt{\pi}}\exp[\Lambda_-(h,h_2)]. \quad \text{(A17)}$$

Substituting the connection: $e^{\Lambda_-(h_1,h)}e^{\Lambda_-(h,h_2)}=e^{\Lambda_-(h_1,h_2)}$ into Eq. (A17), we can find $\alpha_1=2\beta_2e^{\Lambda_-(h_1,h_2)}$ and $\alpha_2=2\beta_1e^{\Lambda_-(h_1,h_2)}$. When $h$>>$h_2$, only the transmitted forward wave exists, so that the WKB solution is given by $\phi_{\rm WKB}(h)=F[L^2(h)]^{-\frac{1}{4}}\times\exp[i\Lambda_+(h_2,h)]$, where $F$ is transmittance. In this region, the WKB solution should equal to the MAF solution. By using the asymbotic form of Airy funtion again, we have

$$\frac{\alpha_2}{\sqrt{\pi L(z)}}\sin\left[\Lambda_+(h_2,h)+\frac{\pi}{4}\right]+\frac{\beta_2}{\sqrt{\pi L(h)}}\cos\left[\Lambda_+(h_2,h)+\frac{\pi}{4}\right]=\frac{F}{\sqrt{L(h)}}\exp[i\Lambda_+(h_2,h)]. \quad \text{(A18)}$$

Eq. (A18) further gives a connection between $\alpha_2$ and $\beta_2$, i.e., $\alpha_2 = i\beta_2, \beta_2 = \sqrt{\pi}\exp\left(-i\frac{\pi}{4}\right)F$ . Now, the three coefficients $\alpha_1$, $\beta_1$, and $\alpha_2$ can be expressed by $\beta_2$ (or $F$), only $F$ needs to be determined.

We consider the initial condition for $h \ll h_1$, then WKB solution can be assumed as

$$\phi_{\mathrm{WKB}}\,(h) = A[L^2(h)]^{-\frac{1}{4}}\exp[i\Lambda_+(h_i, h)] + B[L^2(h)]^{-\frac{1}{4}}\exp[-i\Lambda_+(h_i, h)], \tag{A19}$$

where $A$ and $B$ represent the incident and reflecting coefficients, respectively, and $h_i$ is the incident position of light. Similarly, the WKB solution should equal to the asymbotic form of MAF solution, that is

$$\frac{\alpha_1}{\sqrt{\pi L(h)}}\sin\left[\Lambda_+(h, h_1) + \frac{\pi}{4}\right] + \frac{\beta_1}{\sqrt{\pi L(h)}}\cos\left[\Lambda_+(h, h_1) + \frac{\pi}{4}\right] = \frac{A}{\sqrt{L(h)}}\exp[i\Lambda_+(h_i, h)] + \frac{B}{\sqrt{L(h)}}\exp[-i\Lambda_+(h_i, h)]. \tag{A20}$$

From Eq.(A20), we obtain $\left(\beta_1 - \frac{\alpha_1}{i}\right)e^{-i\left[\Lambda_+(h_i,h_1)+\frac{\pi}{4}\right]} = 2\sqrt{\pi}A$. By replacing $\alpha_1$ and $\beta_1$ with $F$ and assuming $A$=1 we have $F = \frac{e^{i\Lambda_+(h_i,h_1)}}{e^{\Lambda_-(h_1,h_2)}+\frac{1}{4}e^{-\Lambda_-(h_1,h_2)}}$. Thus, all the four coefficients are determined.

**Funding.** National Natural Science Foundation of China (Grants No.62375241).

**Disclosures.** The authors declare no conflicts of interest.

**Data availability.** Data underlying the results presented in this paper are not publicly available at this time but may be obtained from the authors upon reasonable request.